\documentclass{aa}
\usepackage{graphicx}
\usepackage{natbib}
\usepackage{lscape}
\bibpunct{(}{)}{;}{a}{}{,}
\begin{document}

\title{Kinematics and dynamics of the M51-type galaxy pair \\ NGC 3893/96 (KPG 302)}

\author{I. Fuentes-Carrera \inst{1}\thanks{Presently at the
Observatoire de Paris-Meudon}
 M. Rosado
\inst{2} \and P. Amram \inst{3} \and H.Salo \inst{4} \and E.
Laurikainen \inst{4} }

\offprints{I. Fuentes-Carrera}
%, \email{isaura@astro.iag.usp.br}}

\institute{Instituto de Astronom\'\i a, Geof\'\i sica e Ciencias
  Atmosf\'ericas, Universidade de S\~ao Paulo,
Rua do Mat\~ao 1226-Cidade Universit\'aria, 05508-900 S\~ao Paulo SP, Brazil\\
\email{isaura@astro.iag.usp.br,{\bf isaura.fuentes@obspm.fr}}
 \and Instituto de Astronom\' \i a,
Universidad Nacional Aut\'onoma de M\'exico (UNAM),
  Apdo. Postal 70-264, 04510,
M\'exico, D.F., M\'exico \\
\email{margarit@astroscu.unam.mx}
 \and Laboratoire d'Astrophysique
de Marseille, 2 Place Le Verrier,
Marseille Cedex 4, France \\
\email{philippe.amram@oamp.fr}
 \and Department of Physical
Sciences, Division of Astronomy,
University of Oulu, FIN-90570, Oulu, Finland \\
\email{heikki.salo@oulu.fi,eija.laurikainen@oulu.fi} }

\date{Received . . . . . . . . . .  ; accepted . . . . . . . . . . }

\abstract
  {}
% aims heading (mandatory)
{We study the kinematics and dynamics of the M51-type interacting
galaxy pair KPG 302 (NGC 3893/96). We analyse the perturbations
induced by the encounter on each member of the pair, as well as
the distribution of the dark matter (DM) halo of the main galaxy
in order to explore possible differences between DM halos of
"isolated" galaxies and those of galaxies belonging to a pair.
}
% methods heading (mandatory)
{The velocity field of each galaxy was obtained using scanning
Fabry-Perot interferometry. A two-dimensional kinematic and
dynamical analysis of each galaxy and the pair as a whole is done
emphasizing the contribution of circular and non-circular
velocities. Non-circular motions can be traced on the rotation
curves of each galaxy allowing us to differentiate between motions
associated to particular features and motions that reflect the
global mass distribution of the galaxy. For the main galaxy of the
pair, NGC 3893, optical kinematic information is complemented with
HI observations from the literature to build a multi-wavelength
rotation curve. We try to fit this curve with a mass-distribution
model using different DM halos.}
% results heading (mandatory)
{Non-circular motions are detected on the velocity fields of both
galaxies. These motions can be associated to perturbations due to
the encounter and in the case of the main galaxy, to the presence
of structure such as spiral arms.
%The location of the corotation radius of this galaxy is also explored.
We find that the multi-wavelength rotation curve of NGC
3893, "cleaned" from the effect of non-circular motions, cannot be
fitted neither by a pseudo-isothermal nor by a NFW DM halo.}
  % conclusions heading (optional), leave it empty if necessary
   {}

%}
%\authorrunning{Fuentes-Carrera et al}
\titlerunning{Galaxy pair KPG 302}

\maketitle

\section{Introduction}

The difference between the mass distribution implied by the
luminosity of a disc galaxy and the distribution of mass implied
by the rotation velocities offers strong evidence that disc
galaxies are embedded in extended halos of dark matter (Sofue \&
Rubin 2001 and references there-in). Detailed knowledge of dark
matter (DM) haloes around galaxies holds important clues to the
physics of galaxy formation and evolution and is an essential
ingredient for any model aiming to link the observable Universe
with cosmological theories. In practice, realistic DM haloes are
neither static nor spherically symmetric \citep{kneb04} and it is
still unknown if their structure and distribution is intrinsically
related to the environment of their galaxies. The question remains
if there exists an intrinsic difference between the DM halo of an
"isolated" galaxy, the DM halo of a galaxy belonging to a pair or
that of a galaxy that is part of a larger group such as a compact
group or a cluster.

 In this sense, rotation curves (RCs) are a
powerful tool to study the distribution of matter (both baryonic
and non-baryonic) in interacting disc galaxies. For a description
of the classical method for studying mass distribution see
\citet{blais01} and references there-in. RCs also allow us to
determine the maximum rotation velocity of a galaxy and thus infer
the total mass within a certain radius using methods such as that
of \citet{leq83}. Nevertheless, care must be taken when using
kinematic information from interacting galaxies since they are
subject to kinematical perturbations that may affect the correct
determination of a RC that actually traces the global mass
distribution of the galaxy. For this reason, 3D spectroscopy
observations are required to separate circular from non-circular
motions in the velocity field of a galaxy and its rotation curve
as shown in \citet{ifc04}.

In this work, we present scanning Fabry-Perot observations of the
M51-type interacting galaxy pair NGC 3893/96 (KPG 302). Section 2
present the scanning Fabry-Perot (FP) observations and data
reductions. Section 3 introduces the pair of galaxies KPG 302 (NGC
3893/96). In section 4 we present the kinematic information
derived from the F-P observations. Section 5 presents the
dynamical analysis of both galaxies, mass estimates as well as the
mass distribution for NGC 3893. The discussion and conclusions are
presented in sections 6 and 7, respectively.

\section{Observations and Data Reductions}

Observations of NGC 3893/96 (KPG 302) were done at the 2.1m
telescope at the OAN-SPM (M\'exico) using the scanning Fabry-Perot
interferometer PUMA \citep{ros95}.  PUMA is a focal reducer built
at the Instituto de Astronom\'\i a-UNAM used to make direct images
and Fabry-Perot (FP) interferometry of extended emission sources
(field of view $10 \arcmin $). The FP used is an ET-50 (Queensgate
Instruments) with a servostabilization system having a free
spectral range of $19.95 \  \AA $ ($912 \ km \ s^{-1} $) at
$H\alpha$. Its finesse ($ \sim 24$) leads to a sampling spectral
resolution of $0.41 \ \AA \ (19.0\  km \  s^{-1})$ achieved by
scanning the interferometer free spectral range through 48
different channels. A $1024 \times 1024$ Tektronix CCD detector
with a resolution of 0.58 $ \arcsec /  pixel$ was used. We used a
$2 \times 2 $ binning in order to enhance the signal. The final
spatial sampling equals $1.16 \ \arcsec / pixel$. In order to
isolate the redshifted  $ H \alpha \ (\lambda_{at \ rest} =
6562.73 \ \AA ) $ emission of the galaxies, we used an
interference filter centered at  $ 6584 \ \AA $ with a $ FWHM$ of
$ 10 \AA $. To average the sky variations during the exposure, we
got two data cubes with an exposure time of 48 minutes each (60 s
per channel). These data cubes were co-added leading to a total
exposure time of  96 minutes. For the calibration we used a H lamp
whose $6562.78 \ \AA $ line was close to the redshifted nebular
wavelength. Two calibration cubes were obtained at the beginning
and at the end of the galaxy observation to check the metrology.

Data reduction and analysis were done using mainly the
ADHOCw\footnote{{\it http://www.oamp.fr/adhoc/adhocw.htm}
developed by J. Boulesteix} and CIGALE softwares \citep{lec93}.
Standard corrections (cosmic rays removal, bias subtraction,
flat-fielding,...) were done on each cube. Once the object cubes
were co-added, the night sky continuum and $6577.3 \AA$  OH sky
line were subtracted. A spectral Gaussian smoothing ($\sigma = 57
\ km \ s^{-1}$) was also performed. Once the spectral smoothing
was done, the calibration in wavelength was fixed for each profile
at each pixel using the calibration cube. The Fabry-Perot scanning
process allows us to obtain a flux value at pixel level for each
of the 48 scanning steps. The intensity profile found along the
scanning process contains information about the monochromatic
emission ($H\alpha$) and the continuum emission of the object. The
continuum image computation was done considering the mean of the 3
lowest intensities of the 48 channels cube. For the monochromatic
image, the $H \alpha $ line intensity was obtained by integrating
the monochromatic profile in each pixel. The velocity maps were
computed using the barycenter of the $H\alpha$ profile peaks at
each pixel. In order to get a sufficient signal-to-noise ratio on
the outer parts of each galaxy, we performed three spatial
Gaussian smoothings ($\sigma = 2.36, \ 3.54, \ 4.72 \ \arcsec$) on
the resulting calibrated cube. A variable-resolution radial
velocity map was built using high spatial resolution (less
spatially-smoothed pixels) for regions with originally higher
signal-to-noise ratio.

\section{NGC 3893 and NGC 3896}

NGC 3893/96 is an interacting galaxy pair with number 302 in the
Catalogue of Isolated Pairs of Galaxies in the Northern Hemisphere
(KPG, Karachentsev 1972). Morphologically it resembles M51 (NGC
5194/95) since it is composed of a main spiral galaxy and a
considerably smaller companion. KPG 302 is situated in the Ursa
Major cluster. With only 79 members, this cluster is poorly
defined with a velocity dispersion of only $ 148 \  km \ s^{-1}$
and a virial radius of $ 880 \ kpc$ \citep{tul96}. It contains
essentially only late-type galaxies distributed with no particular
concentration toward the center. Given the isolation criteria {\bf
used in the KPG} and the nature of the cluster, it is possible
that the DM halo (or haloes) of the galaxies in the pair are
isolated from those of the cluster.

\begin{figure*}
\centering
\includegraphics[scale=0.62, angle=0] {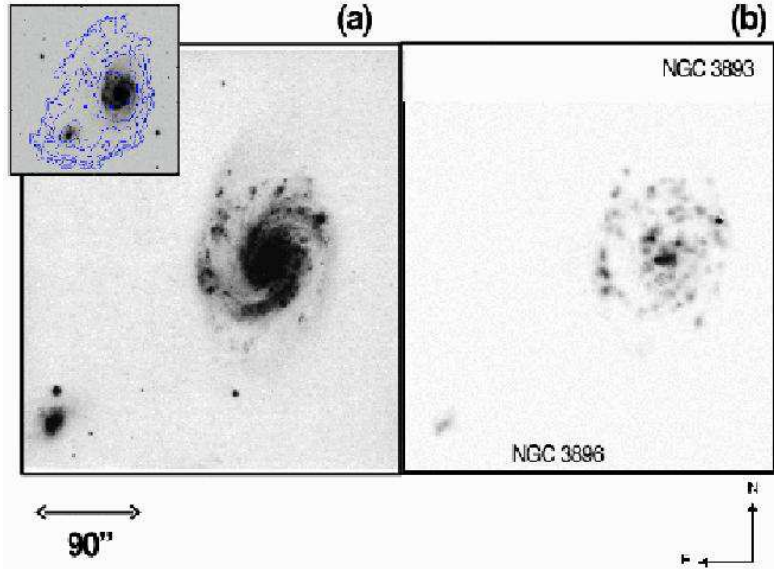}
\caption{ {\it a)} Direct B image  of  NGC 3893/96 (KPG 302) from
``The Carnegie Atlas of Galaxies. Volume II'' \citep{sand94}. {\it
b)} Monochromatic $H \alpha$ (continuum substracted) image  of the
pair obtained from the scanning Fabry-Perot interferometer PUMA
data cubes. {\it Upper panel:} Optical image with HI isophotes
superposed. Image taken from  \citet{ver01} in "An HI Rogues
Gallery" {\it
(http://www.nrao.edu/astrores/HIrogues/webGallery/RoguesGallery06.html)}
} \label{k302_cont_ha}
\end{figure*}

NGC 3893 is a grand-design spiral similar to NGC 5194 in M51
(Figure \ref{k302_cont_ha}a). It has been classified as SABc in
LEDA database, and as SAB(rs)c in NED\footnote{The NASA/IPAC
Extragalactic Database (NED) is operated by the Jet Propulsion
Laboratory, California Institute of Technology, under contract
with the National Aeronautics and Space Administration.} database
and in the RC3 \citep{dvac91}. However, \citet{hec01} classify it
as a non-barred galaxy without any inner ring. HI observations
\citep{ver01}, show that this galaxy is slightly warped in its
outer parts  -both in its HI distribution and its HI kinematics.
Its companion, NGC 3896, appears to be an intermediate type galaxy
between and S0 and a spiral, having also a bar. It shows extended
$H\alpha$ emission. NGC 3893 shows no color excess, whereas NGC
3896 has predominantly blue $B - V$ color in the central parts of
the galaxy \citep{lau98,hec01}.

The star-formation rate (SFR) of each member of the pair was
derived by \citet{jam04}. These authors found a value of $ 5.62 \
M_\odot \ yr^{-1}$ for NGC 3893 and a value of $ 0.14 \ M_\odot \
yr^{-1}$ for its companion. NGC 3893 was also part of a dynamical
analysis of high surface spiral galaxies using long-slit
observations and numerical modelling in order to quantify the
luminous-to-dark matter ratio inside their optical radii
\citep{krz03}. According to these authors, NGC 3893 has a massive
stellar disc which dominates the dynamics of the central regions
with a disc mass of $2.32 \times 10^{10}$ solar masses. They infer
the location of the corotation resonance of this galaxy at $5.5
\pm 0.5 \ kpc$. Though optical images of this pair do not show an
apparent connection between the two galaxies, radio images by
\citet{ver01} show extended HI emission encompassing both galaxies
(small panel in Figure \ref{k302_cont_ha}). This common HI
envelope is elongated from SE to NW, parallel to the line that
joins the nuclei of both galaxies. HI isophotes also show what
could be considered as a broad arm going from NGC 3893 to NGC
3896. Table 1 lists the main parameters of each galaxy.

\section{Kinematic results}

\subsection{Monochromatic Images}

Figure \ref{k302_cont_ha}b displays the monochromatic $H\alpha$
image of the pair. Knotty HII regions lie along the spiral arms of
NGC 3893. Though they mainly follow the two main arms, they also
outline small segments of fainter and less pronounced flocculent
arms.
Intense HII regions are seen on the east side of the galaxy along
the western spiral arm. A very bright HII region is seen in the
central parts of the galaxy.

For NGC 3896, important $H\alpha$ emission is seen within the
inner 13.9 $\arcsec$ (1.3 $kpc$) southeast of the center of the
galaxy. This emission displays two maxima that form a rather
elongated region (left panel in Figure \ref{k302_cont_ha}b). The
northern parts of the galaxy show weak diffuse emission. This type
of emission is also seen on the western side of the galaxy. These
sides are closer to NGC 3893.

\subsection{Velocity Fields}

\begin{figure*}
\centering
 \includegraphics[scale=0.56]{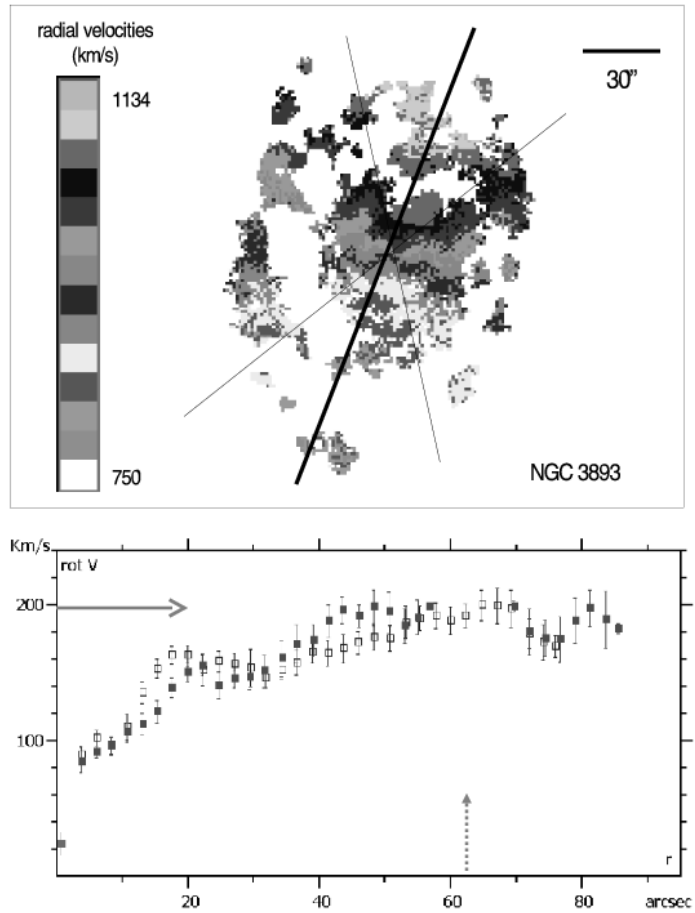}
\caption{{\it Top:} Velocity field of NGC 3893 in KPG 302. Solid
line indicates the
  galaxy's position angle $(P.A.)$,
slash-dotted lines indicate the angular sectors from both sides of
the major axis considered for the computation of the galaxy's the
computation of the galaxy's rotation curve. {\it Bottom:} Rotation
curve (RC)
  of NGC 3893. Both sides of the curve have been superposed. Open squares
  correspond to the receding side of the galaxy. Filled squares correspond to
  the approaching side. RC was plotted considering an inclination value of 45$^\circ$.
  Horizontal solid arrow indicates the maximal rotation velocity.
  Vertical dotted arrow indicates the radius associated with
  corotation according to \citet{krz03}.}
\label{ngc3893_vf}
\end{figure*}

\begin{figure*}
\centering
 \includegraphics[scale=0.49] {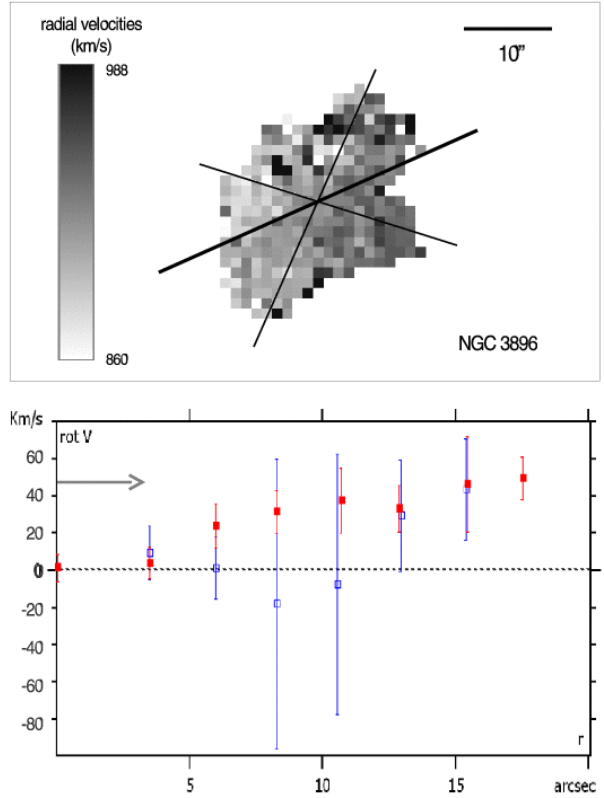}
\caption{{\it Top:} Velocity field of NGC 3896 in KPG302. Solid
line indicates the galaxy's position angle $(P.A.)$,  slash-dotted
lines indicate the angular sectors from both sides of the major
axis considered for the computation of the galaxy's RC.  {\it
Bottom:} RC
  of NGC 3893. Both sides of the curve have been superposed. Open squares
  correspond to the receding side of the galaxy. Filled squares correspond to
  the approaching side. RC was plotted considering an inclination value of 49$^\circ$.
  Solid arrow
indicates maximal rotation
  velocity.}
\label{ngc3896_vf}
\end{figure*}

The upper panel in Figure \ref{ngc3893_vf} shows the velocity
field of NGC 3893. It is a smooth and regular field; the
isovelocities show no significant distortions. Although the galaxy
has an elongated inner structure in the monochromatic image (see
Figure \ref{k302_cont_ha}), the central parts of the galaxy show
no kinematic signatures that could be associated to the presence
of a nuclear bar. The minor kinematic axis is almost perpendicular
to the main axis of the galaxy - outlined by isovelocities with
$955 \ km \ s^{-1}$. Small wiggles seen along these isovelocities
could be a signature of an inner ring, similar to that shown by
the simulations by \citet{salo99} of IC4214. The velocity field is
symmetrical with respect to the kinematical minor axis. Locally,
minor irregularities are seen in the distribution of radial
velocities, especially along the spiral arms of the galaxy. These
might be associated to the passage of gas through the spiral
density wave.
For NGC 3896, the velocity field is very perturbed displaying a
mild velocity gradient from the SE to the NW (upper panel in
Figure \ref{ngc3896_vf}). Isovelocities are very patchy and
crooked.

\subsection{Rotation Curves}
\label{bozomath}

In the case of early-stage interactions, the inner parts of
galaxies are not strongly perturbed, velocity fields are still
smooth and symmetrical resulting in symmetric and low-scattered
rotation curves (RCs) up to a certain radius {\ bf \citep{ifc04}}.
With this assumption in mind, the RC of each galaxy was computed
considering different values for the kinematical parameters
involved in order to obtain a symmetric curve in the inner parts
of the galaxy and to minimize scatter on each side of the curve.
The rotation curves are sampled with bins of 2 pixels ($\sim 2.3
\arcsec$). Error bars give the dispersion of the rotation
velocities computed for all the pixels found inside each
elliptical ring defined by the successive bins. This approach is
described in more detail in \citet{ifc04} and \citet{olive05}.

\subsubsection{NGC 3893}

The RC of NGC 3893 was computed considering  points on the
velocity field within an angular sector of $ 32 ^\circ $  on each
side of the galaxy's  $P.A.$ The kinematic center used to compute
the rotation curve  matches the photometric center from the PUMA
continuum image within $ 1 \arcsec $. The most symmetrical, smooth
and less-scattered
 RC was derived using the following set of values:
 $P.A. =  (340  \pm 10)   ^\circ   $,  $i =  (45 \pm 3)   ^\circ   $ and
 $V_{sys} = (962  \pm 5) \  km \  s^{-1} $. These are presented in Table
 1 along with values obtained in previous works. The RC
 superposing both the approaching and receding sides is shown in the lower
panel in Figure \ref{ngc3893_vf}. Globally the RC of NGC 3893 is
symmetric up to the last $H\alpha$ emission point at $85 \arcsec$
($7.55 \ kpc$). Both sides of the curve display oscillations of
about $10 \ km \ s^{-1} $. The maximum rotation velocity  is  $
(197 \pm 10) \ km \ s^{-1}$ and is reached at $69 \arcsec$ ($6.1 \
kpc$). Our RC is very similar to that derived by \citet{olive02}
using FP observations. In order to compare with the results
derived by \citet{krz03}, a long-slit was simulated by considering
radial velocities on the velocity field within a sector of $ 1.5
^\circ $ on each side of the slit position angle. The values of
the kinematic parameters were equal to those used by these
authors. These "long-slit" RCs show the same increasing behavior
and oscillations as those found by \citet{krz03}.

\subsubsection{NGC 3896}

Though the velocity field of this galaxy is very perturbed, we
were still able to derive a RC reflecting the circular motions of
the southwestern side of the galaxy (approaching side). The
following set of values were used:
 $P.A. =  (294  \pm 5)   ^\circ   $,  $i =  (49 \pm 3)   ^\circ   $ and
 $V_{sys} = (920.50  \pm 5) \  km \  s^{-1} $, considering an angular sector
 of $ 42 ^\circ $  on each side of the galaxy's  $P.A.$. These values along
 with values from previous works are presented in Table 1.
 The lower panel in Figure \ref{ngc3896_vf} shows the RC derived for this galaxy
 superposing both the approaching and
receding sides. Globally the curve displays an increasing behavior
up to the last emission point at $17.5 \arcsec$ ($1.5 \ kpc$)
where the velocity equals $ 48 \ km \ s^{-1}$.
Rather small velocity values appear near the center of the galaxy,
within $ 4.2 \arcsec$ ($ 0.4 \ kpc$).

\subsection{Non-circular motions}
\label{kin_morph}

\begin{figure*}
\centering
\includegraphics[scale=0.62]{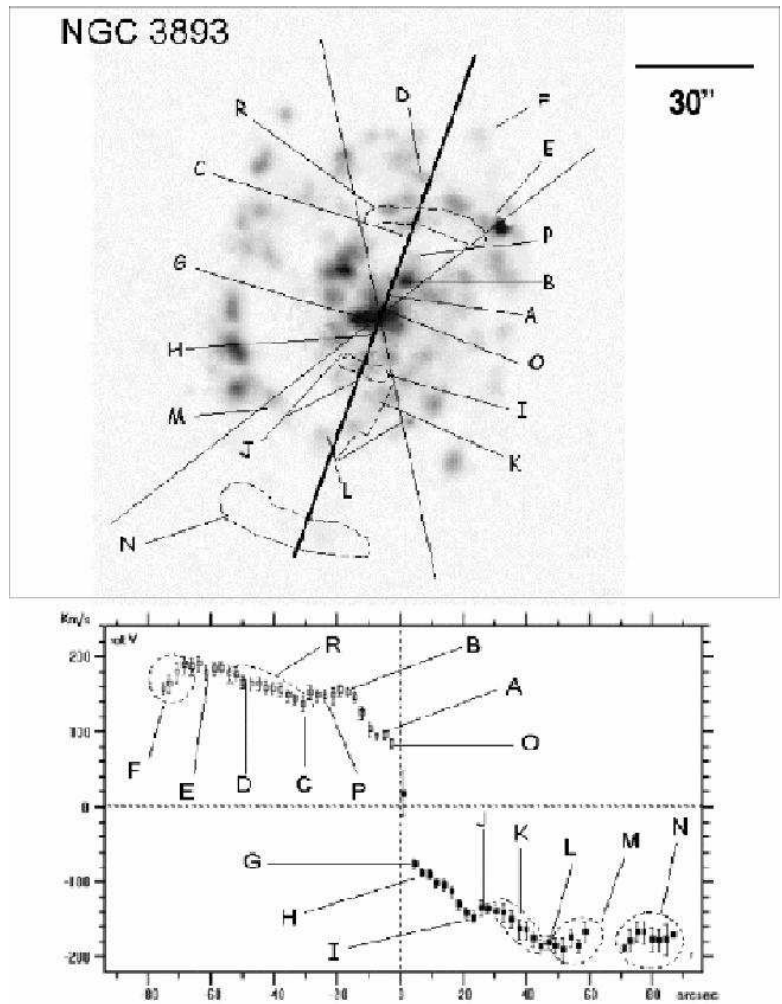}
\caption{Top: Monochromatic image of NGC 3893. Letters indicate
features associated to points on the RC (shown in the bottom
panel) in order to differentiate the contribution of circular from
non-circular motions. Solid line indicates the galaxy's position
angle $(P.A.)$, the thin lines indicate the angular sector from
both sides of the major axis considered for the computation of the
galaxy's RC.}
\label{ngc3893_monolet}
\end{figure*}

Two dimensional kinematic fields of disc galaxies portray the
motion of the gas all over the galaxy enabling us to match these
motions with different morphological structures. One can determine
to which extent the gas is following circular motion around the
center of the galaxy and to which extent there are important
contributions from non-circular velocities (radial, azimuthal and
vertical ) due to the presence of these structures or external
perturbations. For NGC 3893, we analysed the influence of
morphological features on the kinematics of the gas. This was done
by comparing the monochromatic image with  each side of the RC
(Figure \ref{ngc3893_monolet}). Through this comparison we were
able to differentiate points in the RC associated to circular
motions of the gas -associated with the global mass distribution
of the galaxy- from non-circular motions -associated to the
response of the gas to local morphological features.

\section{Dynamical Analysis}

\subsection{Mass Estimates through Rotation Curves}

\begin{figure*}
\centering
\includegraphics[scale=0.62, angle=0]{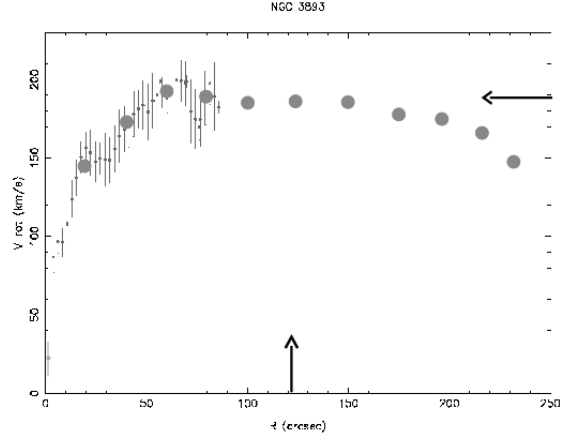}
\caption{Multi-wavelength curve of NGC 3893. Small dots in the inner parts of the curve
  correspond to optical Fabry-Perot $H\alpha$ observations. Larger dots in the
  outer parts correspond to the HI curve derived by \citet{ver01}. The
  horizontal arrow indicates the point with maximum rotation
  velocity. Vertical arrow shows the radius considered for mass estimation
  using the method by \citet{leq83}.}
  \label{ngc3893_multiL}
\end{figure*}

The $H\alpha$ kinematic information from our FP observations of
NGC 3893  was complemented with $HI$ synthesis observations by
\citet{ver01}. In order to match both curves, we considered the
averaged optical rotation curve which was derived using the same
kinematic parameters as those used for the HI RC (these values do
not differ much from the values we used for the optical RC -see
Table 1). The superposed curves are shown in Figure
\ref{ngc3893_multiL}. Both curves superpose smoothly except for
the external parts of the optical RC ($ R \geq 60 \arcsec$) which
appear to be more perturbed. The $HI$ RC also displays a
decreasing behaviour for $ R \geq 175 \arcsec$ and up to the last
$HI$ emission point at $ R = 245 \arcsec$ ($21.76 \ kpc$).
\citet{ver01} mention that the determination of the HI rotation
curve beyond $ 120 \arcsec $ is uncertain because of the tidal
interaction with NGC 3896. We shall set this radius as our limit
for the computation of the mass through the RC of NGC 3893. Using
this composite rotation curve, a range of possible masses was
computed using the method proposed by \citet{leq83} according to
which the total mass of a galaxy within a radius $R$ lies between
$ 0.6$ (in the case of a disc-like mass distribution) and $1.0$ $
\times (RV^2(R) / G) $ (in the case of a spheroidal mass
distribution). As a first estimate, we considered the maximum
rotation velocity ($ 190 \ km \ s^{-1}$) given by both the optical
and radio RCs to estimate the mass within $R = 120 \arcsec = 10.8
\ kpc = 0.99 \ D_{25}/2$. The range of masses within this radius
is given by $0.50$ to $0.84 \times 10^{11} M_\odot$. In order to
derive the mass of NGC 3896, we considered the last emission point
on the approaching side of the curve which shows the maximum
amplitude of that curve. This value equals 48 $km \ s^{-1}$ at $R
= 17.5 \arcsec = 1.5 \ kpc = 0.4 \ D_{25}/2$. The range of masses
within this radius is from $4.78$ to $7.97 \times 10^{8} M_\odot$.
%Mass estimates for both galaxies and their ratio are shown in Table \ref{LES_masses}.
The mass ratio of the galaxies was computed based on the NIR
luminosities, calculated from the $K_s$-magnitudes taken from
2MASS \citep{skrut06}.
% and shown in Table \ref{LES_masses}.
These luminosities for NGC 3893/96 lead to the mass ratio of
0.031. On the other hand, the mass ratio derived from the RC of
each galaxy within 0.4 $D_{25}/2$ falls around 0.0255. These
values are fairly similar, particularly if considering the
uncertainty of the mass-to-light ratio. This fact strengthens the
reliability of the mass estimates based on the rotation curves.
%Table \ref{LES_masses} summarizes these results.

\subsection{Mass Distribution}

In order to study the mass distribution in NGC 3893 we used the
mass model from \citet{blais01}. This model uses both the light
distribution of the galaxy and a theoretical dark halo profile to
compute a RC that best fits the observed one. The mass-to-light
ratio of the disc  $(M/L)_{disk}$) as well as the properties of
the dark matter, characteristic density ($\rho_{0}$) and radius
($R_{0}$), are free parameters. We used a DM halo described by a
pseudo-isothermal sphere \citep{beg87} and a Navarro, Frenk \&
White (NFW) profile \citep{nav96} in order to fit the
multi-wavelength RC of NGC 3893. Optical photometry in the I band
was taken from \citet{hec01} and the $HI$ superficial distribution
from \citet{ver01}. Making use of the possibility to disentangle
circular from non-circular velocities in the optical RC, our
multi-wavelength RC was "cleaned" from points associated to
non-circular motions (see section \ref{kin_morph}) and to the warp
in the outer parts of the HI disc. We removed points in the
optical part of the observed RC between $15 \arcsec$-$20 \arcsec$,
$22 \arcsec$-$32 \arcsec$ and at R$> 75 \arcsec $ (see Figure
\ref{ngc3893_multiL}).
Only points from the HI rotation curve were taken into account
after $75 \arcsec $ and up to $120 \arcsec$ considering the fact
that the RC in HI is uncertain beyond this radius.

\begin{figure*}
\centering
\includegraphics[scale=0.25, angle=270]{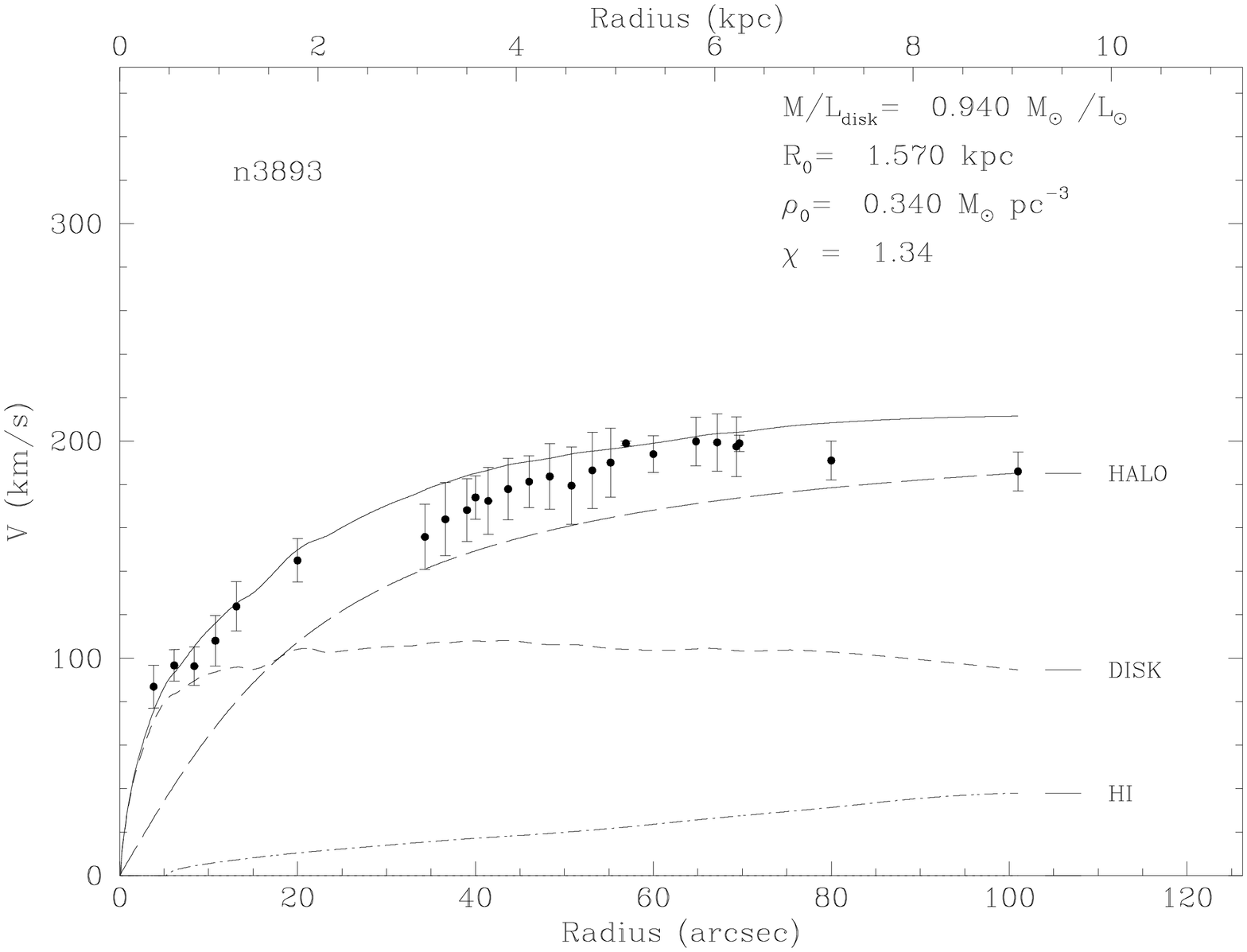}
\includegraphics[scale=0.25, angle=270]{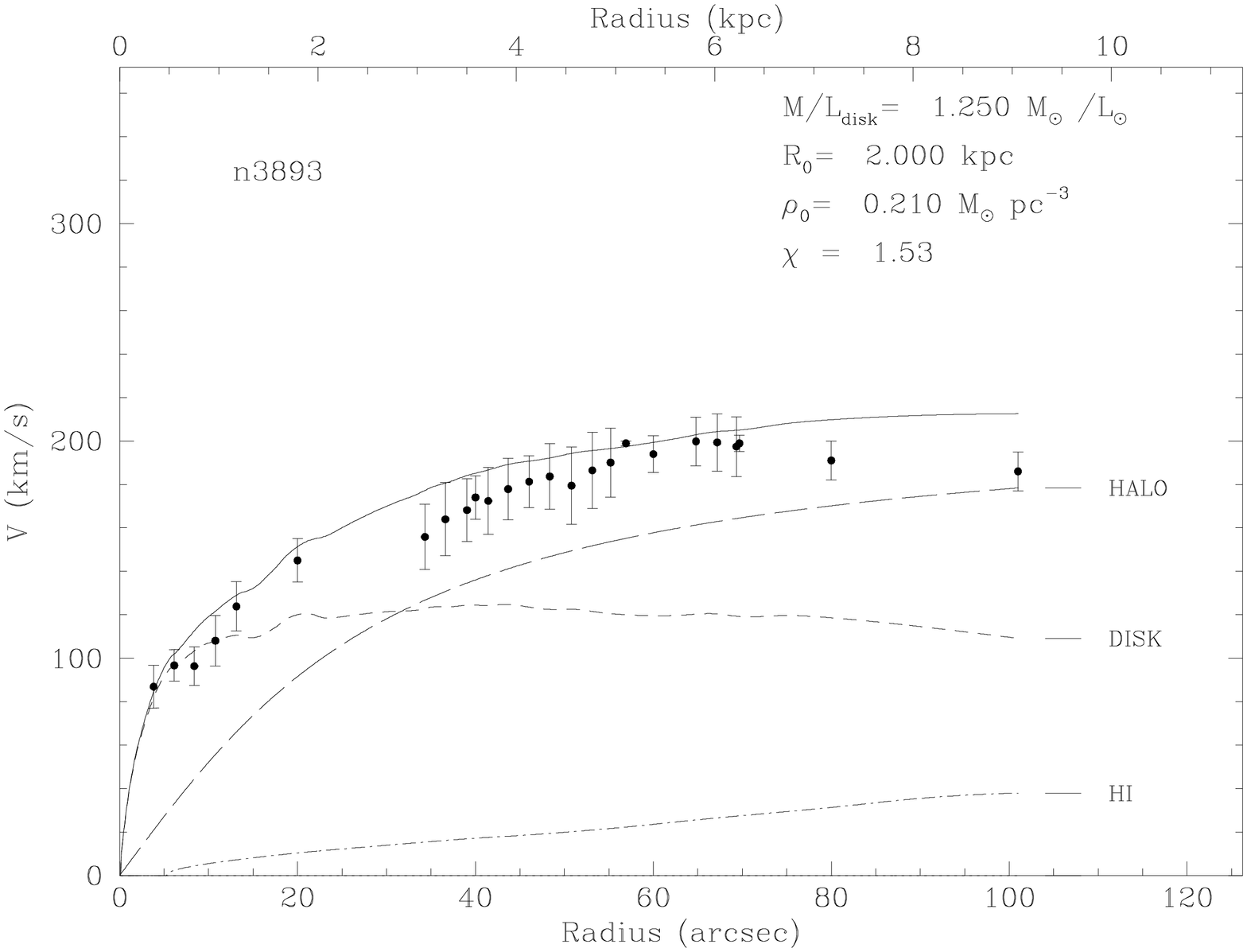}
\includegraphics[scale=0.25, angle=270]{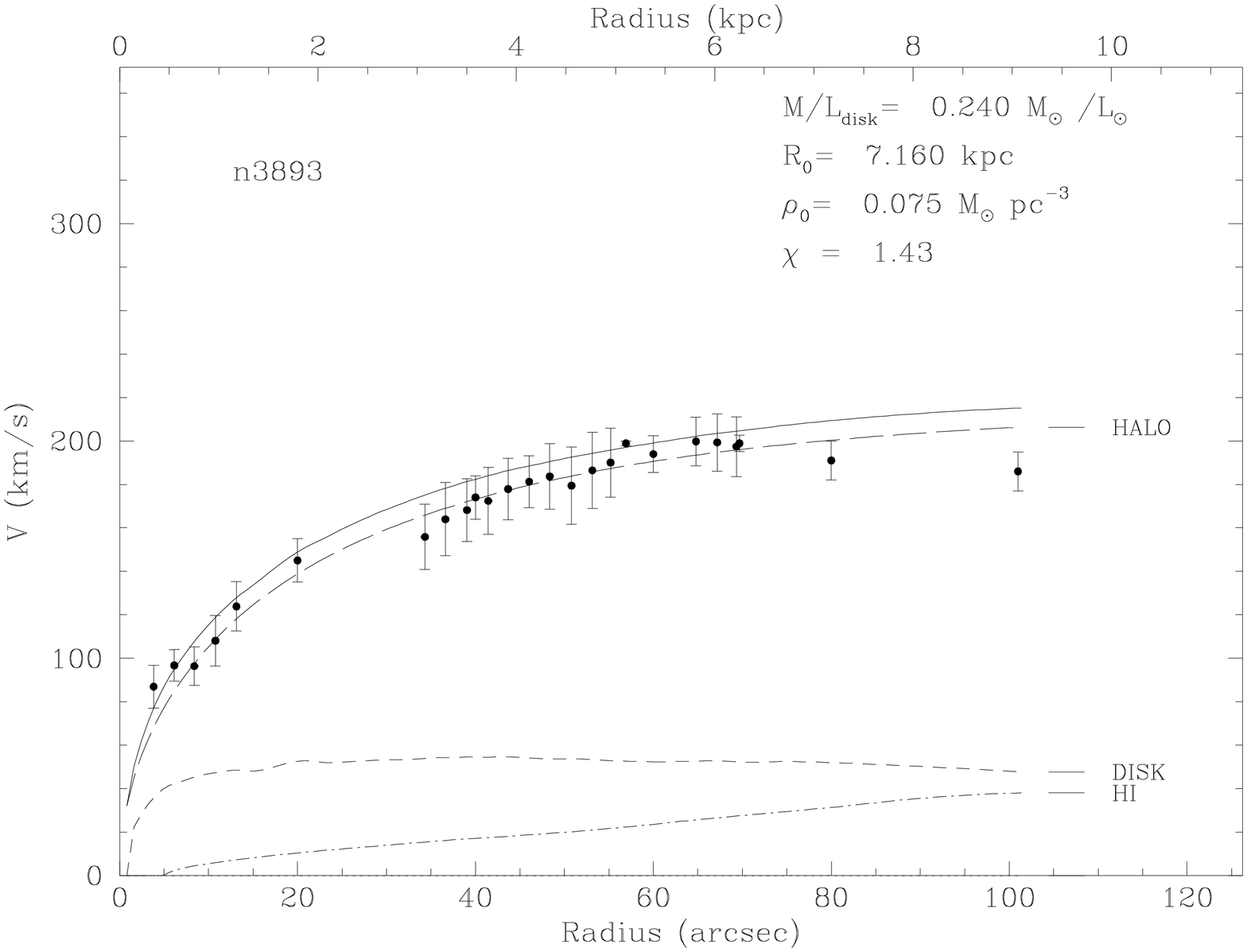}
\includegraphics[scale=0.25, angle=270]{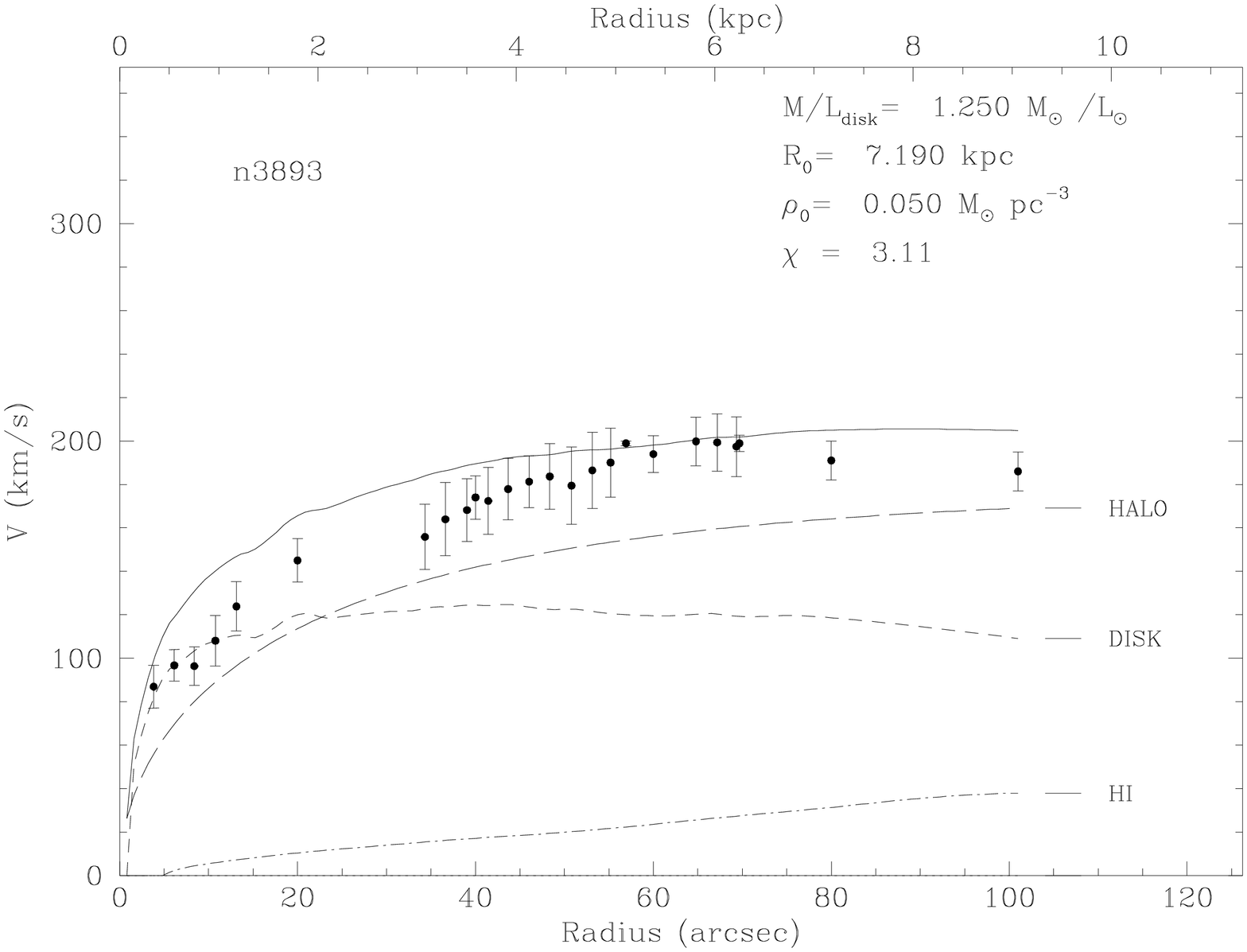}
\caption{Best mass model fits for the multi-wavelength rotation
curve of NGC 3893 once the points associated with non-circular
motions have been removed. {\it Top left:} Pseudo-isothermal halo
and non-maximal disc. {\it Top right:} Pseudo-isothermal halo and
maximal disc. {\it Bottom left:} NFW halo and non-maximal disc.
{\it Bottom right:} NFW halo and maximal disc. Long-dashed curve
represents the dark-matter halo contribution, short-dashed curve
represents the stellar disc contribution. The parameters displayed
stand for the mass-to-light ration of the stellar disc
($M/L_{disc}$), the characteristic radius of the dark matter halo
and  density ($R_0$ and $\rho_0$, respectively) and the minimized
$\chi^2$ in the three-dimensional parameter space. Mass-model
taken from \citet{blais01}.} \label{RCtot_nonpert}
\end{figure*}

Figure \ref{RCtot_nonpert} shows different fits for this
multi-wavelength RC: a pseudo-isothermal DM halo with non-maximal
disc ({\it top left}), a pseudo-isothermal DM halo with a maximal
disc ({\it top right}), a NFW halo with non-maximal disc ({\it
bottom left}) and a NFW halo with a maximal disc ({\it bottom
left}). Table \ref{LE_massmod} displays the mass model parameters
used in each case. We used the definition of \citet{sack97} for
the "maximal disc" which is taken to be a galactic disc such that
$ 85 \% \pm 10\%$ of the total rotational support of a galaxy at a
radius $ 2.2 x scale radius$ is contributed by the stellar disc
mass component. For NGC 3893, this radius corresponds to 1.80 $
kpc$ \citep{krz03}. The best fit ($\chi^2 = 1.34$) is obtained
using a pseudo-isothermal halo with a non-maximal disc leading to
$(M/L)_{disc} = 0.94$
in the $I$ band, yet it misses the last two points of the RC. The
fit with a NFW halo and a non-maximal disc gives $(M/L)_{disc} =
0.24$ ($\chi^2 = 1.43$), also missing the two outermost points of
the RC. Both a pseudo-isothermal and NFW halo with a maximal disc
give larger values of  $(M/L)_{disc}$ ($1.25$
in both cases) and also larger $\chi^2$ values (1.53 and 3.11,
respectively). They also miss both outer and inner points on the
RC.

\begin{figure*}
\centering
\includegraphics[scale=0.25, angle=270]{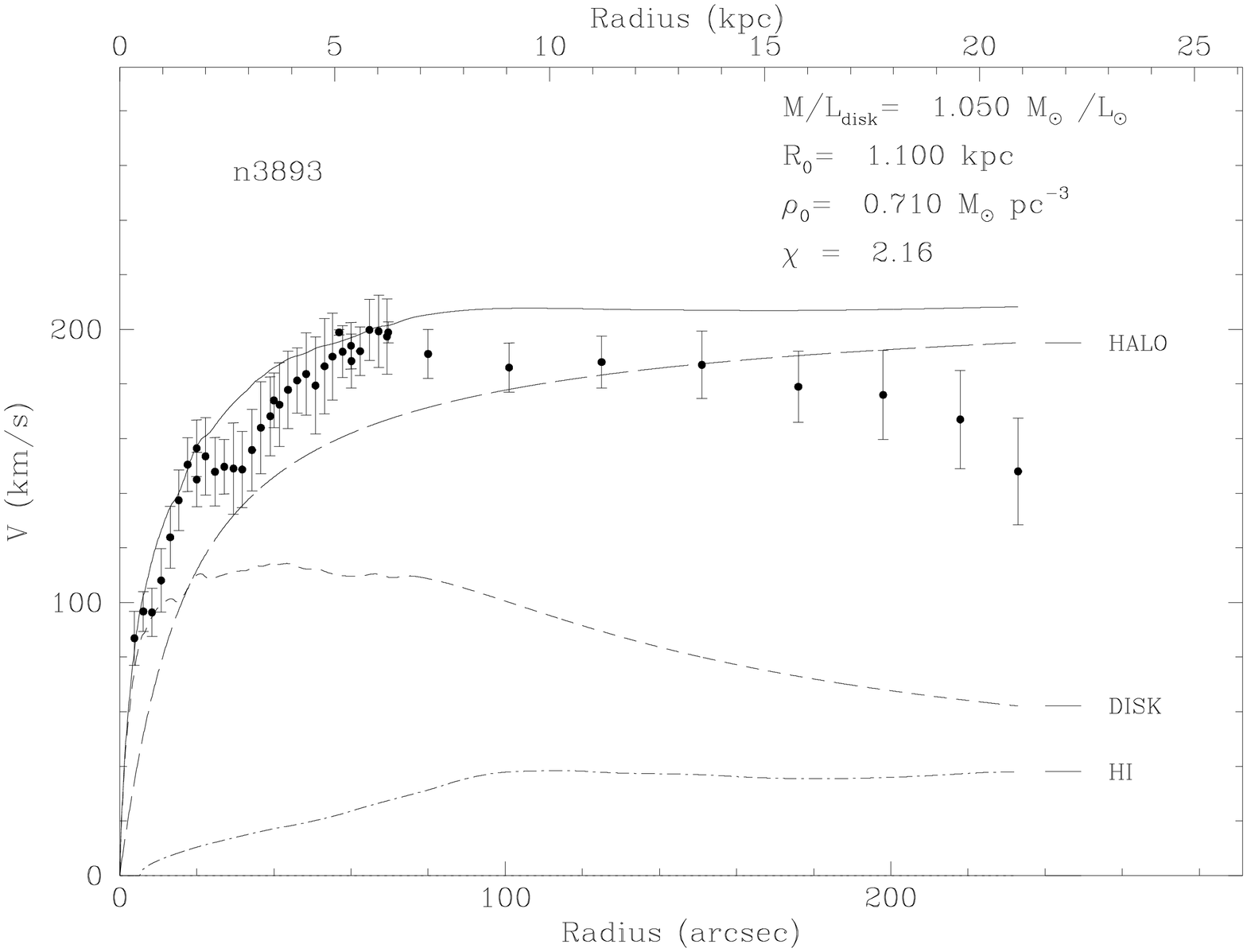}
\includegraphics[scale=0.25, angle=270]{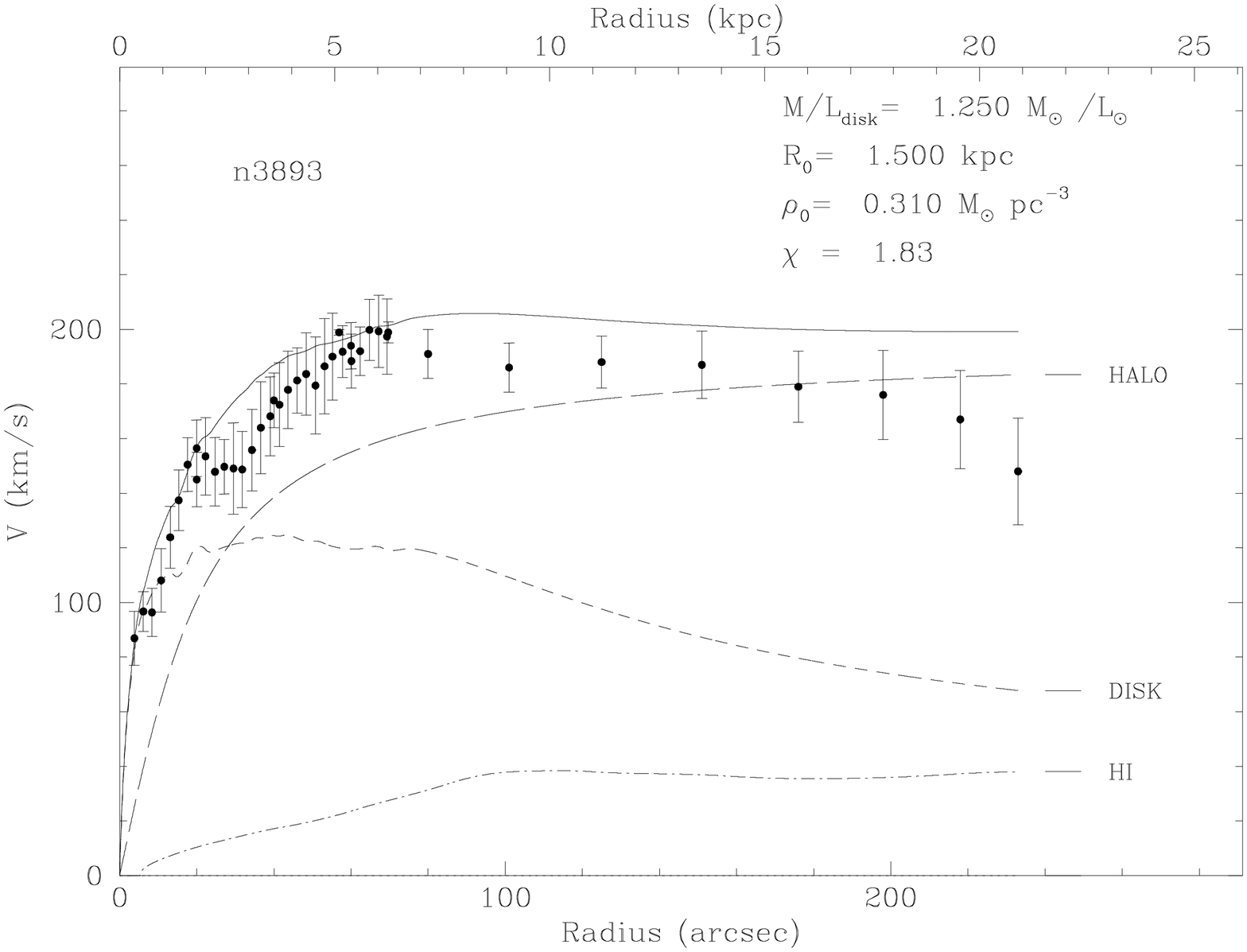}
\includegraphics[scale=0.25, angle=270]{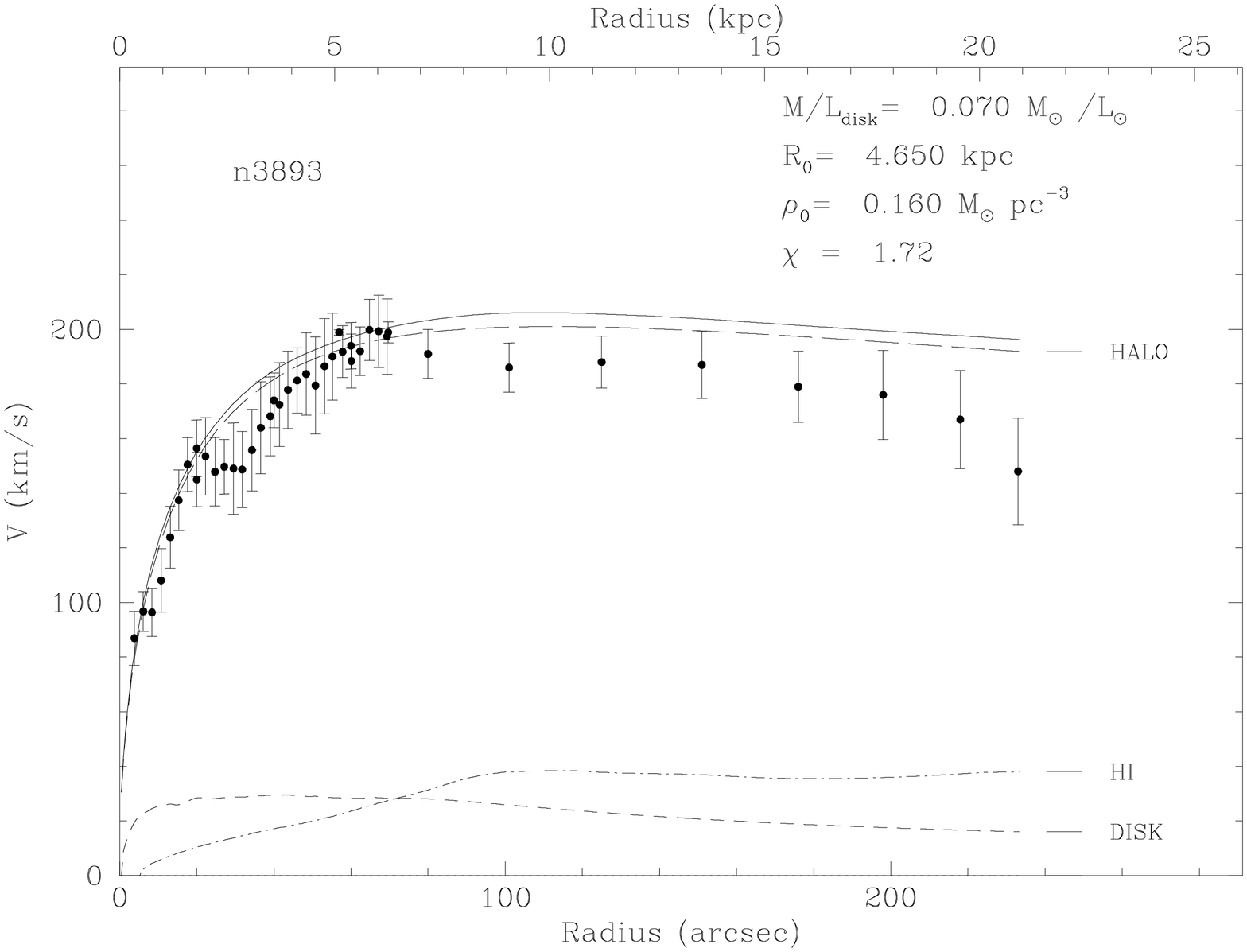}
\includegraphics[scale=0.25, angle=270]{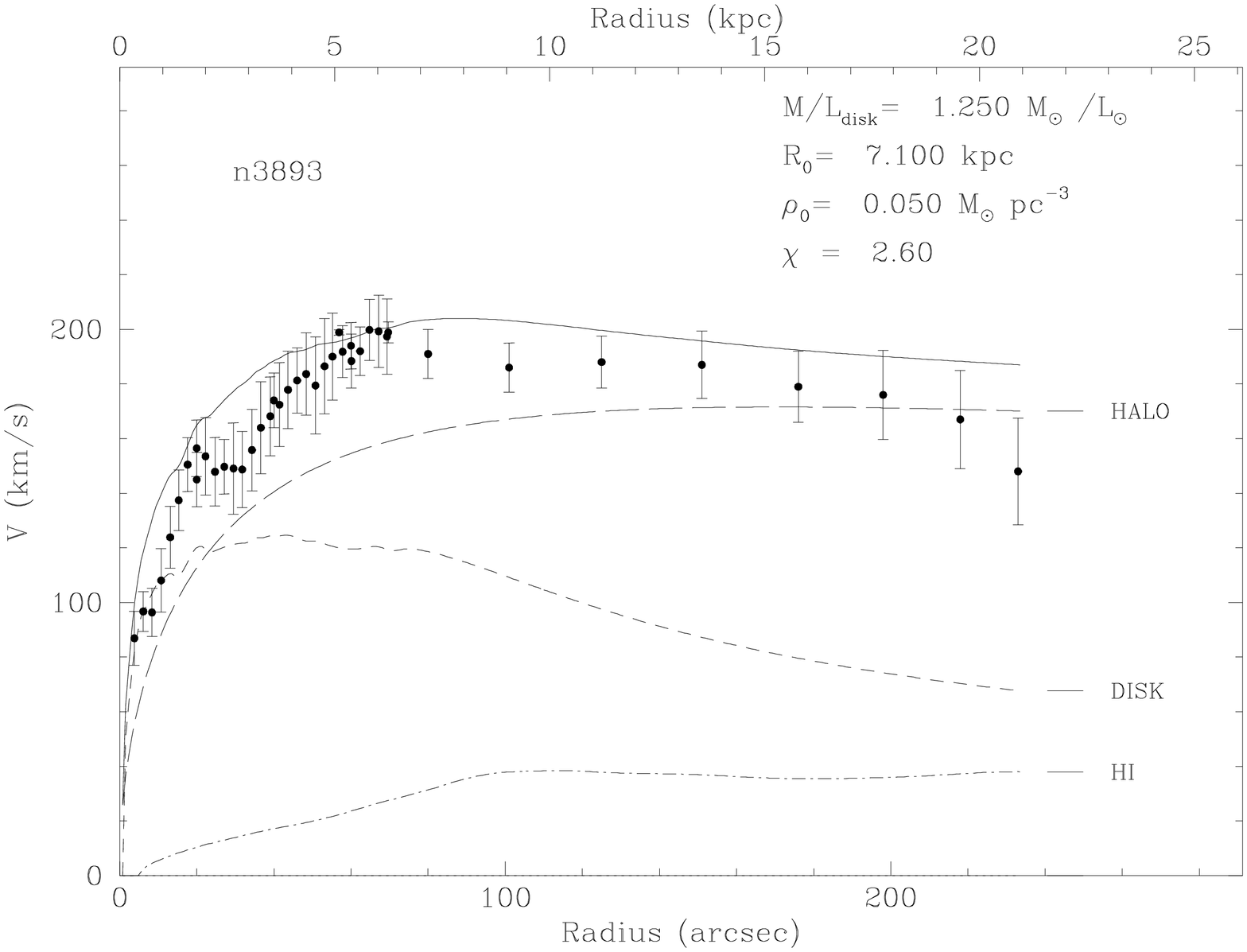}
\caption{Best mass model fit for the multi-wavelength rotation
curve of NGC 3893 considering all observed points -including those
associated with non-circular motions. {\it Top left:}
Pseudo-isothermal halo and non-maximal disc. {\it Top right:}
Pseudo-isothermal halo and maximal disc. {\it Bottom left:} NFW
halo and non-maximal disc. {\it Bottom right:} NFW halo and
maximal disc.}
\label{RCtot_all}
\end{figure*}

\begin{figure*}
\centering
\includegraphics[scale=0.18, angle=270]{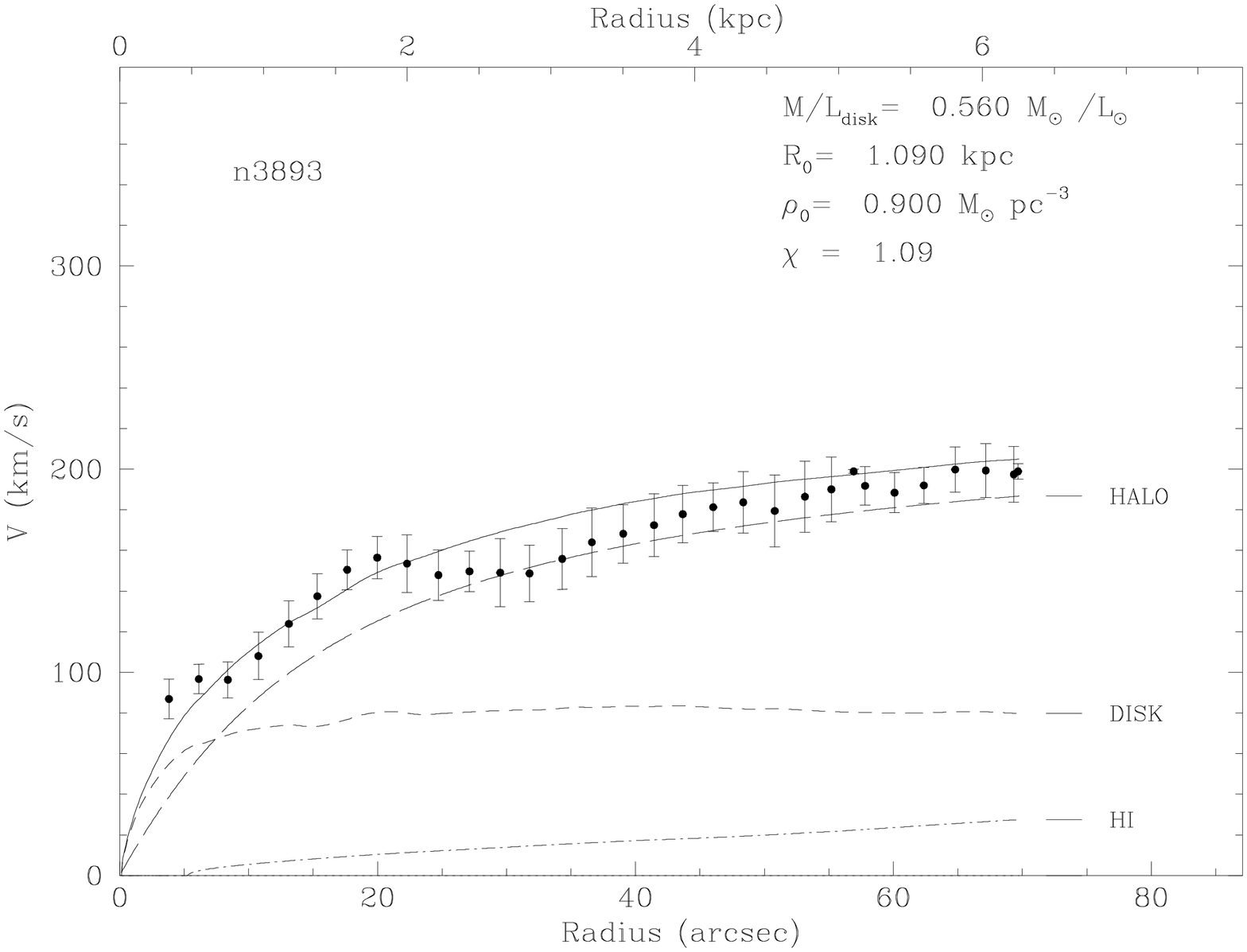}
\includegraphics[scale=0.18, angle=270]{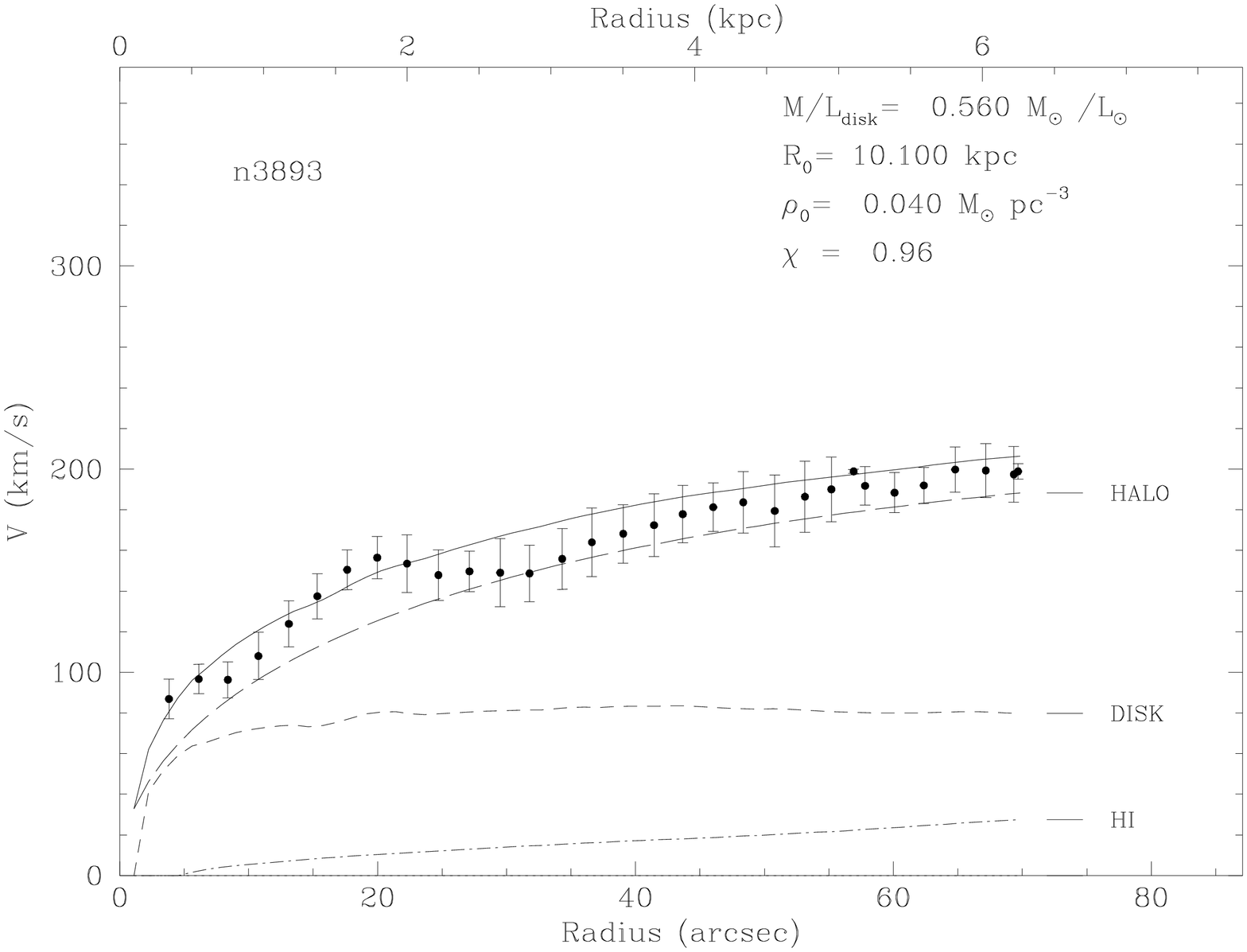}
\caption{Best mass model fit for the H$\alpha$ rotation curve of
NGC 3893 considering all observed points -including those
associated with non-circular motions and using the $ (M\L)_{disc}
$ value by \citet{krz03} . {\it Left:} Pseudo-isothermal halo.
{\it Right:} NFW halo. }
\label{RC_Ha}
\end{figure*}

\begin{figure*}
\centering
\includegraphics[scale=0.18, angle=270]{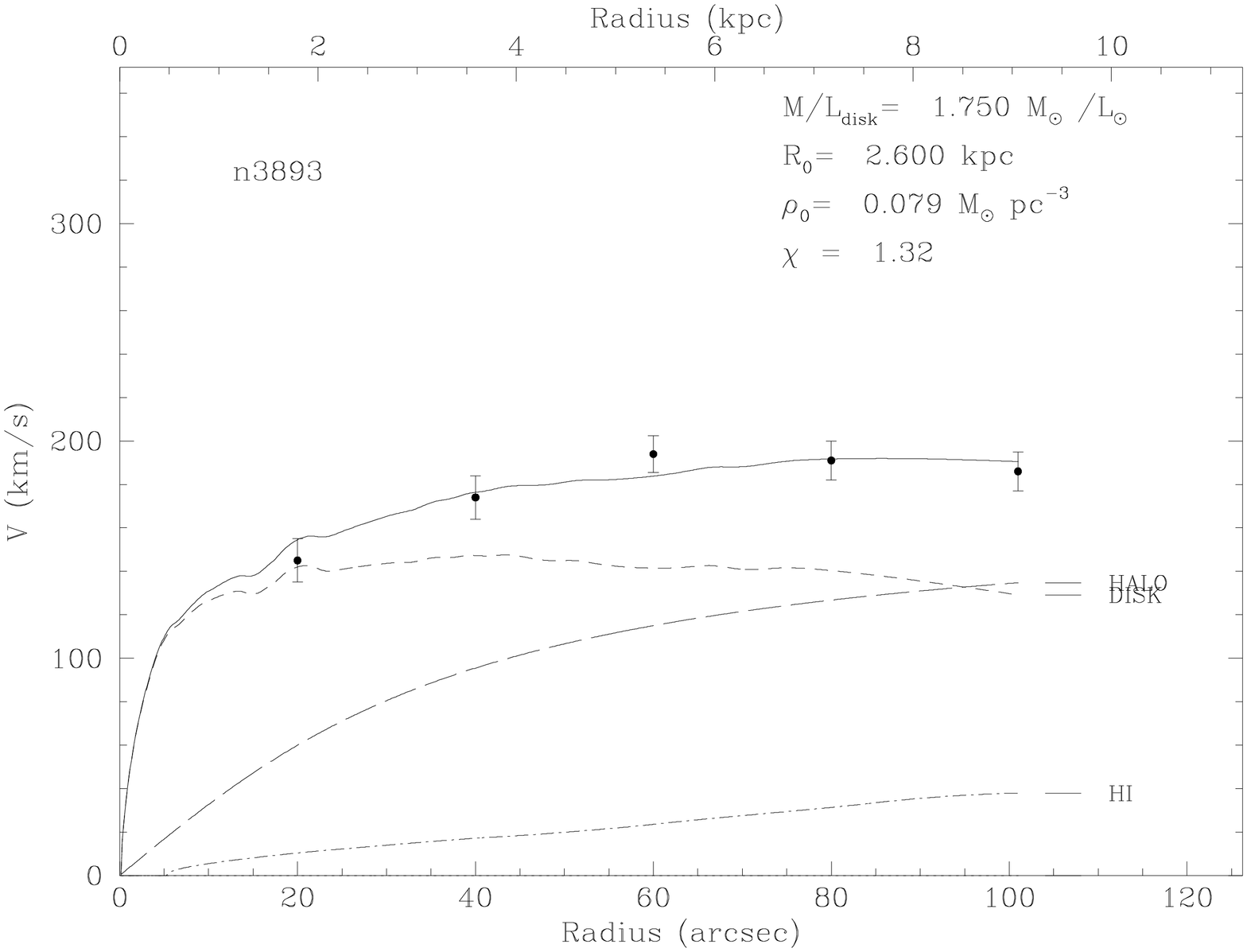}
\includegraphics[scale=0.18, angle=270]{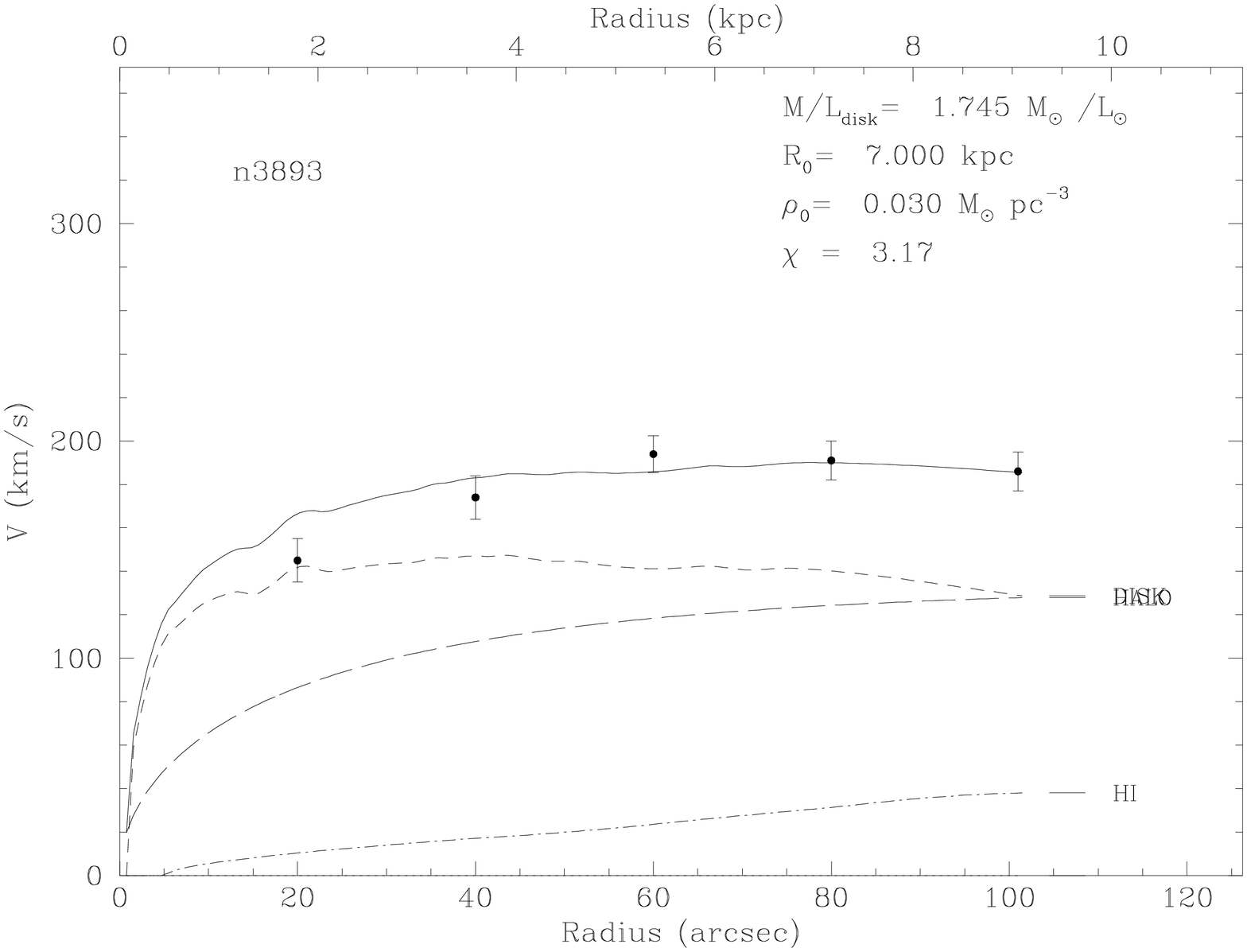}
\caption{Best mass model fit for the HI rotation curve of NGC 3893
after removing points associated with non-circular motions and the
warp of the outer parts of the disc. {\it Left:} Pseudo-isothermal
halo. {\it Right:} NFW halo. } \label{RC_HI}
\end{figure*}

In order to evaluate the effect of non-circular motions on the
fits, we fitted the above mass models including the points
associated to non-circular motions to the multi-wavelength RC.
The fits can be seen in Figure \ref{RCtot_all}. The
values for the mass model parameters are shown in Table
\ref{LE_massmod}. These fits are less precise.
In all cases, the mass-to-luminosity ratio using the maximal disc
assumption is larger than that obtained by \citet{krz03}
($(M/L)_{disc} = 0.56$ -in both K and I bands). We must take into
account the fact that \citet{krz03} only fit the optical part of
the RC derived with long-slit spectroscopy. For the sake of
comparison, we fitted our H$\alpha$ RC (with and without points
associated to non-circular motions) using the value for
$(M/L)_{disc}$ derived by these authors. Results are shown in
Figure \ref{RC_Ha} and Table \ref{LE_massmod}. Fits are very good
for both the pseudo-isothermal and the NFW halo, nevertheless
$(M/L)_{disc} = 0.56$ does not render the disc maximal. Finally we
also fitted these mass models to the HI RC without points at $R
> 120 \arcsec$ where tidal effects might affect the correct
determination of the RC. Figure \ref{RC_HI} and Table
\ref{LE_massmod} show the resulting parameters of the fits. The
fit using the NFW halo misses the inner most point of the curve
and is rather inaccurate ($ \chi^2 = 3.17$). The fit using the
pseudo-isothermal halo misses the middle point of the curve, yet $
\chi^2 = 1.32$ -which is on the of the lowest values found for all
fits presented. Nevertheless it should be noticed that for both
haloes, the $(M/L)_{disc} > 1.7$ which is larger than the value
found with the multi-wavelength RC. This highlights the importance
of the multi-wavelength approach for the mass model.

\section{Discussion}

The fact that no model fits the last point in the multi-wavelength
RC of NGC 3893 "cleaned" from the effects of non-circular motions
can be explained either by a truncated halo for this galaxy or by
the existence of a common halo for both galaxies. Since the
derived mass of the NGC 3896 is small, then it most probably
resides inside the halo of NGC 3893 and both galaxies share a
single halo. Nevertheless this halo would have a different
distribution than that of an isolated galaxy. When considering an
isothermal halo and the $HI$ curve, the $(M/L)_{disk}$ that best
fits this curve is much higher than the  $M/L$ found for the
multi-wavelength curve. In general, the information on the inner
parts of the galaxy given by the optical observations imposes an
$(M/L)_{disk} \le 1$ which would imply the presence of a disc with
an important population of young stars -which is not the case
given the  $B - V$ value derived by \cite{hec01}. This supports
the idea that the structure of the DM halo of this pair differs
from that of a single disc galaxy.
As shown by \citet{lau01}, in general M51-type pairs companions
have extremely large bulge sizes relative to their disc
scale-lengths. Consequently, the bulge-to-disc luminosity ratios
for the companions were also generally larger than known for any
of the Hubble types of normal galaxy. This is the case for NGC
3896 whose RC displays almost solid body rotation up to the last
emission point.

\section{Conclusions}

We have presented the kinematic and dynamical analysis of the
M51-type galaxy pair, KPG 302 (NGC 3893/96). NGC 3893 is a
grand-design spiral with a regular velocity field that displays no
major distortions. The companion, NGC 3896 displays on-going star
formation which was probably triggered by the interaction with the
main galaxy. This galaxy displays important non-circular motions
in localized regions, especially on the side of the galaxy which
is closer to the companion. The total mass of each galaxy was
derived from the RC. The optical RC of NGC 3893 was matched with
the existing HI curve in order to determine the distribution of
the luminous and dark matter. This multi-wavelength rotation curve
was analysed in the light of the 3D observations in order to
differentiate the contribution of non-circular motions, associated
to particular features, and the contribution of circular motions,
which reflect the mass distribution of the galaxy. No "classical"
DM halo fits the observed rotation curve which could imply a
different mass distribution for the DM halo of M51-type binary
galaxies.

\begin{acknowledgements}
We wish to thank the staff of the Observatorio Astron\'omico
Naconal (OAN-SPM) for their support during PUMA data acquisition.
We also thank C. Carignan for letting us use his mass model.
I.F.-C. thanks the financial support of FAPESP and CONACYT grants
no.03/01625-2 and no.121551 -respectively. M.R. acknowledges
financial support from grants 46054-F from CONACYT and IN100606
from DGAPA-UNAM. H.S. and E.L. acknowledge the support from the
Academy of Finland. We acknowledge the usage of the HyperLeda
database (http://leda.univ-lyon1.fr), the NASA/IPAC Extragalactic
Database (NED) and the Two Micron All Sky Survey.
\end{acknowledgements}

\bibliographystyle{aa}

\begin{flushleft}
\begin{table*}
\centering
\caption[]{Parameters of NGC 3893 and NGC 3896}
\label{glxs_parameters}
\begin{tabular}{l c c c c}
\hline\hline
 \noalign{\smallskip}
              & \ \ \ \  &    NGC 3893  & \ \ \ \ \ \ \ \ \ \ \ &  NGC 3896  \\
\noalign{\smallskip} \hline
 \noalign{\smallskip}
 Coordinates  (J2000)$^{\mathrm{a}}$   &  &     $\alpha = 11h \ 48m \ 38.38  $    &   &  $\alpha = 11h \ 48m \ 56.42s  $   \\
                   &    &      $\delta = +48^\circ \ 42\arcmin \ 34.4\arcsec    $    &    & $\delta = +48 ^\circ \ 40\arcmin \ 29.2\arcsec   $  \\

 Morphological type  &  &  SABc$^{\mathrm{a}}$      &  &  S0-a$^{\mathrm{a}}$ \\

                     &  &  SAB(rs)c$^{\mathrm{b,c}}$      &  &  S0/a$^{\mathrm{c}}$: pec$^{\mathrm{b}}$ \\

                     &  &  Sc$^{\mathrm{d}}$       &  &  SBbc pec$^{\mathrm{d}}$ \\

 $m_{B}$$^{\mathrm{d}}$ (mag) & &  $11.13$  & &  $14.05$ \\

 $B \ - \ V$$^{\mathrm{d}}$ (in mag)   & &$0.56$ & &$0.46$  \\

 $D_{25}   /   2$$^{\mathrm{a}}$ ($\arcmin $)& &  $2.03$ & & $0.75 $ \\

 Distance$^{\mathrm{e}}$ (Mpc)  & & $18.6$    &  & $18.6$ \\

 SFR$^{\mathrm{f}}$ $(M_\odot/yr)$ & & $5.62$    &  & $0.14$ \\

     Heliocentric systemic velocity \ \ $( km/s) $ & &$969 \pm 3$$^{\mathrm{a,g}}$ &  &$959 \pm 12$$^{\mathrm{a,g}}$ \\

                                      & &$973$$^{\mathrm{j}}$ &  &  - - -  \\

                                      & &$962.0 \pm 5$$^{\mathrm{h}}$ &  &$920 \pm 5 $$^{\mathrm{h}}$ \\

                                      & &$958.0$$^{\mathrm{k}}$ &  & - - -\\

         $ V  _{rotmax}$ \ \  $(km/s)$& & $251.2$$^{\mathrm{a,g}}$ &  &  $150$$^{\mathrm{a,g}}$  \\

                                          & & $195$$^{\mathrm{j}}$ &  &   $confused$$^{\mathrm{j}}$   \\

                                          & & $220 \pm 3$$^{\mathrm{l}}$  & &   - - -  \\

                                          & & $197 \pm 10$$^{\mathrm{h}}$ &  &  $50  \pm 10$$^{\mathrm{h}}$  \\

                                      & &$207$$^{\mathrm{k}}$ &  & - - -\\

              P.A. $( ^\circ )$  & &$352$$^{\mathrm{j}}$   &  & - - -  \\

                                & &$166 + 180 $$^{\mathrm{m}}$  &  & - - - \\

                                & &$340 \pm 10 $$^{\mathrm{h}}$   &  & $294 \pm 5 $$^{\mathrm{h}}$ \\

                                      & &$167 + 180 $$^{\mathrm{k}}$ &  & - - -\\

           Inclination  $( ^\circ )$ & &  $49 \pm 2$$^{\mathrm{i}} $  & & $48 \pm 3 $$^{\mathrm{i}} $   \\

                                     & &  $42$$^{\mathrm{m}} $  & &  - - -  \\

                                     & &  $45 \pm 3$$^{\mathrm{h}} $  &  &  $49 \pm 3$$^{\mathrm{h}} $ \\

                                      & &$49$$^{\mathrm{k}}$ &  & - - -\\

   $m_K$$^{\mathrm{n}}$ (mag) &  & $7.891$  & & $11.648$ \\

\noalign{\smallskip} \hline
\end{tabular}

\begin{list}{}{}
\item[$^{\mathrm{a}}$] LEDA database
\item[$^{\mathrm{b}}$] NED database
\item[$^{\mathrm{c}}$] RC3 (de Vaucouleurs et al.(1991)
\item[$^{\mathrm{d}}$] Hern\'andez-Toledo \& Puerari (2001)
\item[$^{\mathrm{e}}$] Tully \& Pierce (2000)
\item[$^{\mathrm{f}}$] James {\it{et al}} (2004), total measured $H\alpha$ +
  [NII] line flux corrected for [NII] contamination
\item[$^{\mathrm{g}}$] From HI observations
\item[$^{\mathrm{h}}$] This work
\item[$^{\mathrm{i}}$] Verheijen (2001), HI observations
\item[$^{\mathrm{j}}$] Verheijen \& Sancisi (2001), HI observations
\item[$^{\mathrm{k}}$] Garrido et al. (2002)
\item[$^{\mathrm{l}}$] Tully et al.(1996)
\item[$^{\mathrm{m}}$] Kranz, Slyz \& Rix (2003)
\item[$^{\mathrm{n}}$] 2MASS (Skrutskie et al. 2006) $K_{ext}$  value taken from the NED database
\end{list}
   \end{table*}
\end{flushleft}

\begin{flushleft}
\begin{table*}
\centering \caption{Mass models parameters for NGC 3893 from best
fits of the multi-wavelength rotation curve considering only
points associated to circular motions} \label{LE_massmod}
 \begin{tabular} {c c c c c c c c c c c c}
\hline\hline
 \noalign{\smallskip}
 & Type of & & Maximal & & $(M/L)_{disc}$ & & $R_0$ & & $ \rho_0$ & & $\chi^2$ \\
 & halo   & &  disc   & & (I band) & & (kpc) & & ($M_\odot/pc^{-3}$) & &  \\
\hline
\multicolumn{12}{l}{\it Multi-wavelength curve, non-circular motions excluded} \\
 & p-ISO & & no  & & 0.940 & & 1.570 & & 0.340 & & 1.34 \\
 & p-ISO & & yes & & 1.250 & & 2.000 & & 0.210 & & 1.53 \\
 & NFW   & & no  & & 0.240 & & 7.160 & & 0.075 & & 1.43 \\
 & NFW   & & yes & & 1.250 & & 7.100 & & 0.050 & & 2.60 \\
\hline
\multicolumn{12}{l}{\it Multi-wavelength curve, non-circular motions included} \\
%\hline
 & p-ISO & & no  & & 1.050 & & 1.100 & & 0.710 & & 2.16 \\
 & p-ISO & & yes & & 1.250 & & 1.500 & & 0.310 & & 1.83 \\
 &   NFW & & no  & & 0.070 & & 4.650 & & 0.160 & & 1.72 \\
 &   NFW & & yes & & 1.250 & & 7.100 & & 0.050 & & 2.60 \\
\hline
\multicolumn{12}{l}{\it H$\alpha$ curve, non-circular motions included} \\
%\hline
 & p-ISO & & no & & 0.560 & &  1.090 & & 0.900 & & 1.09 \\
 &   NFW & & no & & 0.560 & & 10.100 & & 0.040 & & 0.96 \\
\hline
\multicolumn{12}{l}{\it HI curve, non-circular motions excluded} \\
 & p-ISO & & yes & & 1.750 & & 2.600 & & 0.079 & & 1.32 \\
 &   NFW & & yes & & 1.745 & & 7.000 & & 0.030 & & 3.17 \\
  \hline
\end{tabular}
\end{table*}
\end{flushleft}

\end{document}